\documentclass[showpacs,preprintnumbers]{revtex4}
\usepackage{amssymb}
\usepackage{amsmath}
\usepackage{graphicx}

\begin{document}

\date{\today }
\title{ Excitation of Meinel and first negative band \\
system at the collision of electron and proton with nitrogen molecule }
\author{Malkhaz R. Gochitashvili$^{1}$, Roman Ya. Kezerashvili$^{2}$, and
Ramaz A. Lomsadze$^{1}$}
\affiliation{\mbox{$^{1}$Physics Department, Tbilisi State University} \\
Tbilisi, 380028, Republic of Georgia \\
\mbox{$^{2} $Physics Department, New York
City College of Technology, the
City University of New York,} \\
Brooklyn, NY 11201, USA}

\begin{abstract}
The absolute cross sections for the $e-$N$_{2}$ and $p-$N$_{2}$ collisions
for the first negative B$^{2}\Sigma _{u}^{+}-$X$^{2}\Sigma _{g}^{+}$ and
Meinel A$^{2}\Pi _{u}-$X$^{2}\Sigma _{g}^{+}$ bands have been measured in
the energy region of 400$-$1500 eV for electrons and 0.4$-$10 keV for
protons, respectively. Measurements are performed in the visible spectral
region of 400-800 nm by an optical spectroscopy method. The ratio of the
cross sections of the Meinel band system to the cross section of the first
negative band system $(0,0)$ does not depend on the incident electron
energy. The population of vibrational levels corresponding to A$^{2}\Pi _{u}$
states are consistent with Franck-Condon principle. The ratio of the cross
sections of $(4,1)$ to $(3,0)$ bands, and $(5,2)$ to $(3,0)$ bands exhibits
slight dependence on the proton energy. The theoretical estimation within
the quasimolecular approximation provides a reasonable description of the
total cross section for the first negative band.

\vspace{0.1cm} 

\end{abstract}

\maketitle



{}



\section{Introduction}

\label{intro}

\bigskip\ Solar matter impacting the atmosphere is ejected from the Sun
primarily as a result of solar flares, solar wind, coronal mass ejections,
and solar prominences. Solar flares are sporadic events and are caused by
the magnetic instabilities and disturbances in the corona that happen when
energy stored in twisted magnetic fields is suddenly released. The other
source of electrons and protons is the low-energy solar wind which results
from the high temperature of the corona. The solar wind carries an
electron-proton plasma, constantly evaporated by the sun and which is hot
enough to escape the Sun's gravity and to flow outward into space,
continuously shedding electrons and protons. Coronal mass ejection from the
Sun consists of plasma comprised primarily of electrons and protons as well
as small quantities of nuclei such as helium, oxygen, and iron. Solar
prominences are loops of radiant gas ejected from an active region on the
solar surface that move through the inner parts of the corona under the
influence of the Sun's magnetic field. The ionized gas generated by these
phenomena follows the solar magnetic field lines away from the Sun. Most of
it subsequently cools and falls back to the photosphere. Therefore, besides
the electromagnetic radiation, the basic corpuscular parts of the solar
radiation heading towards the Earth that subsequently interact with the
atmosphere are electrons and protons in the energy spectrum between a few
tenth of eV to hundreds of MeV \cite{Roman JBIS}, \cite{Roma 2 New}. They
are deflected by the Earth's magnetic field towards the Poles and get
scattered and absorbed by atmospheric atoms and molecules. The accompanying
ionization of various gases in the upper atmosphere is causing beautiful
displays known as the Aurora - a luminous glow of the upper atmosphere.

While there have been investigations of main characteristics of the Aurora
and their dependence on factors such as altitude, geographic placement,
solar activity etc., there is still a lack of the quantitative description
of this phenomenon. Upon reaching the denser layers of the atmosphere,
electrons and protons participate in various inelastic processes such as
ionization, molecular excitation, and charge-exchange reactions on
atmospheric gases, especially on nitrogen molecules. Spectral analysis of
the Aurora shows that the ionized nitrogen molecules can radiate in the
visible, infrared and ultraviolet region. Usually, the appearance of this
radiation is observed at an altitude of approximately 110 km.

Observations of the prominent first negative ($B^{2}\Sigma
_{u}^{+}-X^{2}\Sigma _{g}^{+}$ ) and Meinel ($A^{2}\Pi _{u}-X^{2}\Sigma
_{g}^{+}$) band system in the ionized nitrogen molecule $N_{2}^{+}$ indicate
their presence in the Aurora and the dayglow \cite{Ramazi 1} - \cite{Ramazi
6}. The $A^{2}\Pi _{u}-X^{2}\Sigma _{g}^{+}$ system of the $N_{2}^{+}$ has
been extensively studied since Meinel \cite{Meinel 50} first observed the
existence of the $A^{2}\Pi _{u}$ excited state in auroral emissions in the
near-infrared and the $A^{2}\Pi _{u}-X^{2}\Sigma _{g}^{+}$ system of the $%
N_{2}^{+}$ became known as the Meinel system.

Because $\ $the $A^{2}\Pi _{u}$ state of the $N_{2}^{+}$ ion is effectively
created during the ionization of the $N_{2}$ molecule by electron impact or
capture of an electron by a proton (hydrogen emissions are the signature of
proton aurora), the spectrum of the Aurora is characterized by the bright
lighting of the Meinel band. These bands appear in the spectra of polar
auroras and carry information on the collision processes which take place
between molecules and electrons in the upper atmosphere. During $e-N_{2}$
and $p-N_{2}$ collisions, a vibrationally excited N$_{2}^{+}$ molecules and
their radiative decay are accompanied by creation of ground electronic X$%
^{2}\Sigma _{g}^{+}$ states. Hence, in the Aurora the relative vibrational
population of the X$^{2}\Sigma _{g}^{+}$ state is partly governed by
electron excitation and relaxation processes of the A$^{2}\Pi _{u}$ states.

\bigskip The study of the emission spectra gives the opportunity to
determine the concentration and energy distribution of particles entering
the upper layer of the atmosphere. To solve this problem, it is necessary to
determine the absolute cross sections of various inelastic processes, such
as ionization, excitation, and charge-exchange with high precision.
Determination of the absolute cross section of the Meinel band system is
especially problematic. Number of experimental works in which the Meinel
band system is measured is very limited, and they are usually related to the
processes of excitation of the Meinel band through electron collisions with
nitrogen molecules \cite{Malx 9} - \cite{Malxaz 15}. In addition, in the
case of electron impact, the experimentally determined cross section for a
formation of the N$_{2}^{+}$ ions in the A-state is also known within 50\%
because measurements of the excitation cross section are connected with
various difficulties. In particular, the lifetime of the nitrogen molecule
ions in the A$^{2}\Pi _{u}$ state is about $10^{-5}$ s \cite{Malx 9, Malx 10}%
, and during measurements the quenching of excited particles (the transfer
of the excited energy to other particles) is expected to occur. There is
only one study \cite{Malxaz 17} of the excitation of the Meinel band $(2,0)$
by a proton impact on the N$_{2}$ molecule in the energy range $1.5-4.5$ keV.

In this article, we present the results of measurements for the electron and
proton impact excitation of the N$_{2}^{+}$ first negative (1NG) and Meinel
band system obtained with a high precision. The measurements are performed
in the visible ($600-800$ nm) spectral region using the optical spectroscopy
method for electron energies 400 -1500 eV and proton energies 0.4 - 10 keV.
A relative population of vibrational states corresponding to the Meinel band
is measured at a sufficiently low pressure ($0.1-0.7$ mTorr). Therefore, the
quenching and excitation effects caused by collisions involving secondary
electrons are minimized. We have measured the absolute value of the cross
sections for the first negative nitrogen molecule ion ( B$^{2}\Sigma
_{u}^{+} $ - X$^{2}\Sigma _{g}^{+}$) with band $(0,0)$ and for the Meinel
system (A$^{2}\Pi _{u}$--X$^{2}\Sigma _{g}^{+}$) with bands $(3,0)$, $(4,1)$%
, and $(5,2)$. The discrepancies in absolute cross sections and theoretical
interpretations are discussed.

The paper is organized in the following way. In Sec. II the research method
is described. In Sec. III the experimental results and their discussion are
presented. Finally, conclusions follow in Sec. IV.

\section{\protect\bigskip\ Research method}

We have used an experimental setup and method of measurements similar to
those described and used in our previous papers \cite{Malxaz 18} - \cite{Mal
new 1}. The protons extracted from the high frequency discharge source were
accelerated, collimated and focused, and mass-selected with a $60^{0}$
magnetic sector field. Then the proton beam was directed into the collision
chamber. In order to make the measurements for the case of electron
collisions we used the electron gun placed into the mass-analyser chamber.
The electron beam was deflected at $90^{0}$ and, after the collimation and
additional focusing, was directed into the collision chamber. The
fluorescence emitted as a result of the excitation of colliding particles
was observed at 90$^{0}$ with respect to the beam. In the present work,
measurements are performed by the optical spectroscopy method that allowed
us to have the sufficiently high energy resolution of $0.001$\ eV and
therefore sufficient to distinguish the excitation channels. This is one of
the advantages of this method, with respect to the collisional spectroscopy
method.\textit{\ }This method also allowed us to estimate the polarization
of excitation, that itself is a powerful tool for establishing the mechanism
for inelastic processes. The spectroscopic analysis of the emission was
performed with a monochromator incorporating a diffraction grating with a
resolution of 40 nm/mm operated in the visible ($400-800$ nm) spectral
region. A polarizer and a mica quarter-wave phase plate are placed in front
of the entrance slit of the monochromator and the linear polarization of the
emission is analyzed. For cancellation of the polarizing effect of the
monochromator, the phase plate was placed after the polarizer and was
rigidly coupled to it. The emission was recorded by a photomultiplier with a
cooled cathode and operated in the current mode.

Calibration of the spectral sensitivity was performed by a tungsten filament
standard lamp, which was chosen due to the lack of reliable experimental
data in the infrared region (the bright Meinel system), that could be used
for the calibration of the system for registration of radiation.\emph{\ }To
obtain the proton beam we use a high frequency (20 MHz) ion source. The
measurements for low energy collisions required a precise determination of
energy of ions or electrons as well as their energy dispersion. To avoid
errors in the measurements of energy of the incident protons and electrons,
we employed the retarding potential method and used the electrostatic
analyzer with a resolution of 500. Thus, the energy resolution of the
electron beam was the same that for protons, i.e. 500. The energy of the
proton and electron beams was calibrated by measuring their energy.\textbf{\ 
}To estimate the dispersion of energy provided by the high frequency ion
source and electron gun we measured the energy of impacting particles. The
results of the measurements for protons accelerated by the potential of 600
V are given in Fig. 1. The measurements show that the energy deviation from
600 eV is approximately 35 eV. This is related \ to the specification of the
high frequency ion source and the selection of ions from plasma. From Fig.
1, we can see that maximum is observed at 566 eV with a half-width about 20
eV. In our experiments the energy dispersion of the protons does not exceed
this value. For electrons the energy dispersion is less then 5 eV.

The proton current in the collision chamber was of the order $1-10$ $\mu $A
while the electron current was $5-20$ $\mu $A. The system was pumped by an
oil-diffusion pump. The operating pressure of the gas under investigation
did not exceed $6\times 10^{-4}$ Torr, so that multiple collisions could be
ignored. The residual-gas pressure did not exceed \ $10^{-7}$ Torr. The
absolute accuracy of the measurements was 30\%. The accuracy of measurements
is related to the following factors: the accuracy of pressure measurements
in the collision chamber; precise and immediate determination of the primary
beam current that produce the radiation that collected by optical system
from the region where the collision occurs; accuracies of the relative and
absolute calibration procedures.

\emph{\ }The precision of\ measurements\ of\ the Meinel\ band\ system\
strongly\ depend\ on\ two\ factors\ -\ quenching\ of\ the excited\ N$%
_{2}^{+} $\ molecule\ in\ the A$^{2}\Pi _{u}$\ state\ and\ overlapping\ of\
Meinel\ bands\ with\ the other\ molecular\ emission\ \cite{Malxaz 7}, \cite%
{Malxaz 8}.\ Most\ of\ the\ Meinel\ bands\ are\ spectrally\ overlapped\
with\ nitrogen\ first\ positive\ emissions \cite{Malxaz 8}, \cite{Malxaz 22}%
, \cite{Mal new 2}. The\ degree\ of\ overlapping\ depends\ on\ pressure\
and\ the\ Meinel\ bands\ are\ less\ overlapped\ at\ low\ pressures\ where\
the\ excitation\ of\ the\ Meinel\ bands\ is\ stronger\ relative\ to\ the\
excitation\ of\ the\ first\ positive\ bands.\ At\ higher\ pressures\ the\
Meinel\ bands\ are\ quenched\ more\ efficiently.\ Also,\ the\ flux\ of\
secondary\ electrons\ increases\ with\ pressure\ and\ the\ secondary\
electrons\ excite\ the\ first\ positive\ bands\ more\ efficiently\ than\
the\ primary\ electrons\ do \cite{Mal new 2}.

To explore an electron excitation quenching effects, we investigated the
spectrum of the Meinel band system in processes of the excitation of the
nitrogen molecule by the electron impact. The experiment was performed with
various densities of the nitrogen target particles. To control excitation
quenching effect, the ratio of the excitation cross section of the Meinel
band and the excitation cross section of the first negative band system ($%
0,0 $) of a nitrogen molecular ion have been measured in various
experimental conditions by varying the concentration of target particles. At
the lowest pressures (less than about 1 mTorr), when the number of created
secondary electrons was small the degree of the first positive band
excitation relative to the Meinel band excitation was at most only a few
percent \cite{Malxaz 8}. The single collision condition was checked by a
linear dependence of the intensity of spectral lines versus target gas
pressure and density of the electron current. Measurements are performed in
the visible ($400-800$ nm) excitation region. To demonstrate a reliability
of the emission spectrum measurements, as an example in Fig. 2, the spectral
scan is shown from 700 $-$715 nm in the first order of the diffraction
grating for the N$_{2}^{+}$ $(A^{2}\Pi _{u})$ $(4,1)$ band \ for the
electron-N$_{2}$ collision at 400 eV. It is clear that there is no influence
due to the collision quenching effect, because the lifetime of the zero
vibrational level of the $B^{2}\Sigma _{u}^{+}$ state is about three orders
of magnitude less than the lifetime of the vibrational level of$\ A^{2}\Pi
_{u}$ states. The optimal experimental conditions were established based on
these considerations.

To obtain the absolute value of the cross section, we use the cross section $%
8.66\times 10^{-17}$ cm$^{2}$ of the first negative system $(0,0)$ for a
proton-N$_{2}$ collision at the energy $7$ keV from Ref. \cite{Malxaz 21} as
reference. In Ref. \cite{Malxaz 21} the analysis of all experimental data of
the exitation state for (391.4 nm) (0, 0) band spectrum in a wide energy
interval is performed. Their analysis shows that the data obtained at a high
energy interval are more reliable (discrepancy in data at energy of protons $%
E>5$ keV is about $\pm $15\%).\emph{\ }It should be mentioned that during
the experiment it was possible to measure the band spectrum for protons and
electrons in the same experimental conditions. This provides an opportunity
to obtain the absolute cross section for the electron collision based on the
proton data. \ For the same experimental conditions, the radiation band for
the first negative system $(0,0)$ was measured for the proton and electron
at energies of 7 keV and of 600 eV respectively. Comparison determined that
the cross section for electrons for the 1NG $(0,0)$ band at the given energy
is $6.6\times 10^{-18}$ cm$^{2}.$

\begin{figure}[t]
\centering
\includegraphics[width = 4.0in]{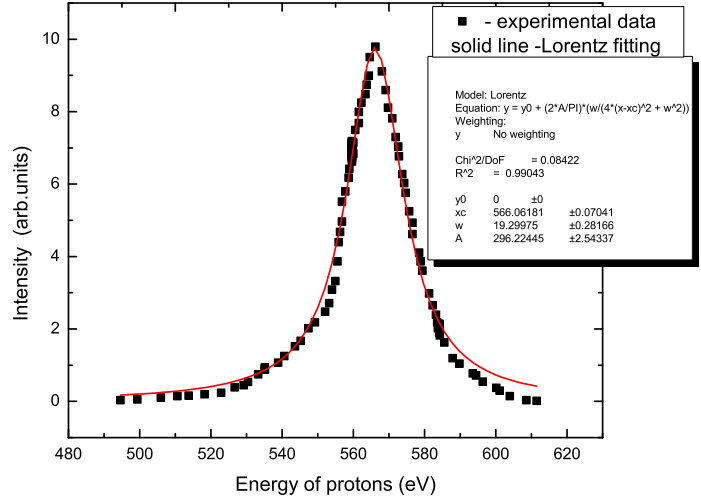}
\caption{Dispersion of the proton energy for the acceleration potential of
600 V.}
\label{lam ht1}
\end{figure}

\begin{figure}[t]
\centering
\includegraphics[width = 3.5in]{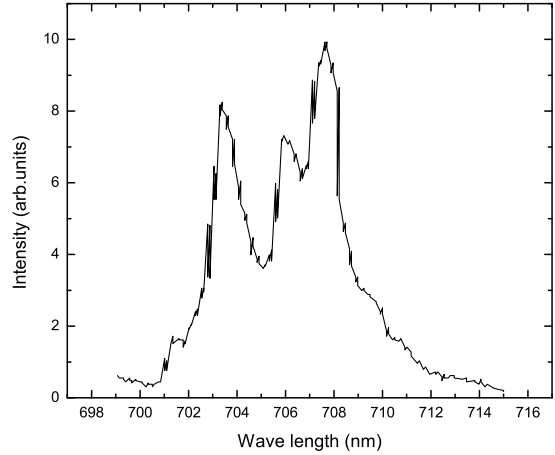}
\caption{Experimental spectrum of the N$_{2}^{+}$ Meinel (4,1) band, for
electron energy of 400 eV.}
\label{lam ht2}
\end{figure}

\section{Experimental results and discussion}

\subsection{Excitation cross sections of N$_{2}^{+}$ by electron impact.}

Firstly, let us present the results for the electron impact excitation of
the N$_{2}^{+}$ nitrogen ion for the first negative and Meinel bands. The
results of the measurements of the absolute excitation cross sections for
the first negative $(0,0)$ and Meinel $(3,0)$, $(4,1)$ and $(5,2$) band
systems by electron impact on N$_{2}$ in the energy region of $400-1500$ eV
are shown in Fig. 3. Analysis of the results shows that the above mentioned
vibrational population of molecule states N$_{2}^{+}$ (A$^{2}\Pi _{u})$ obey
the Franck-Condon principle \cite{Malxaz 231, Malxaz 241}. From the
comparison of the results of our measurements of the cross section for the
1NG $(0,0)$ band system with the data from Ref. \cite{Malxaz 22} shown in
Fig. 3, we can conclude that their magnitude as well as energy dependence
are in good agreement. Note that in Ref. \cite{Malxaz 22} the detail
analysis of all existing data for the excitation of the (0,0) band of the
first negative system by electrons impact was carried out. It was emphasized
that the energy dependence of cross section measured in Ref. \cite{Mal new
29} can be used as a representative in the wide energy interval (19-1000eV).
The recommended data from \cite{Malxaz 22} was obtained by renormalizing the
value of cross section from Ref. \cite{Mal new 29} to the best values
determined in Ref. \cite{Mal new 30} at 100 eV. From Fig. 3, it is clearly
seen that the ratio of the cross sections of the Meinel band system to that
of the first negative band system ($391.4$ nm) $(0,0)$ does not depend on
the energy of incident particles. The same behavior has been obtained by
Piper and coauthors \cite{Malxaz 8}. However, the measured excitation cross
sections of the Meinel band system by the electron impact differ by the
factor of two from the existing measurements \cite{Mal new 31} and \cite{Mal
new 32}.

\begin{figure}[t]
\centering
\includegraphics[width = 3.5in]{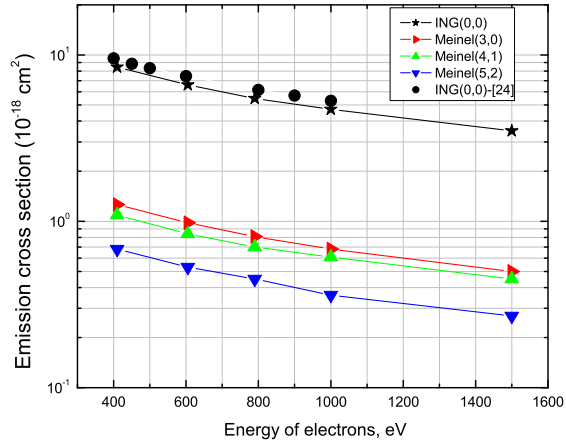}
\caption{Dependence of the excitation cross sections for the first negative $%
(0,0)$ and Meinel $(3,0)$, $(4,0)$, and $(5,2)$ band systems on the electron
energy for the e$-$N$_{2}$ collision.}
\label{lam ht3}
\end{figure}

\subsection{\protect\bigskip Excitation cross sections of N$_{2}^{+}$ by
proton impact.}

To study the excitation of the Meinel band system by proton impact on the
nitrogen molecule, with the omission of the quenching effect, the impulse
gained by the target particle in charge-exchange processes is taken into
consideration. For the case in which the collision processes occur at a
small impact parameter, a large momentum transfer to the target particle
takes place. Therefore, it is expected that the target particles will have
large velocity and, due to large lifetime ($10^{-5}$ s$)$, the excited
particles will escape from the observable excitation area without being
registered by the detector. For this reason, we analyze the results obtained
in Ref. \cite{Malxaz 25} for the differential cross sections of
charge-exchange processes. The data analysis shows, that the scattering
cross section has maximum at very small angles and then at the angle about
of 1 degree it reduced by three orders of magnitude \cite{Malxaz 25}. For
this reason for collision of 10 keV protons the measurements were taken
within the scattered angle limited to $1^{0}.$\ Therefore the estimated the
maximal value of energy obtained by the target particles is determined as%
\textbf{\ }  \cite{Malxaz 26}:

\begin{equation}
\Delta E=\frac{m_{p}}{m_{N_{2}}}E\theta ^{2}f(E\theta )=\frac{1}{28}%
10^{4}\left( \frac{1}{57.3}\right) ^{2}f(E\theta )<0.11\text{ eV,}
\label{eq:1}
\end{equation}%
where $m_{p}$ and $m_{N_{2}}$ are the masses of a proton and N$_{2}$
molecule , respectively, $E$ and $\theta $ are the energy and scattering
angle of the proton, respectively, and the function $f(E\theta )$ reaches
the maximum value of 1 for the elastic limit only. According to Eq. (1) the
velocity of target particles is $v$=10$^{5}$ cm/sec. The particle in the
excited A$-$states with this velocity can run 1 cm distance. In order to
register all of the radiation it is necessary that the height of the
entrance slit of the monochromator would be more than 1 cm and an observed
object in the collision area should be away from the entrance slit at a
distance, that is sufficiently greater than 1 cm. In this case, the
saturation of the excitation along the beam will be kept in advance. During
measurements, the pressure of the target particles was kept sufficiently low
in order to minimize collisional quenching effect. Results obtained for
these conditions are shown in Fig. 4, where the total excitation cross
sections for \ the excitation of the $(3,0)$, $(4,1)$ and $(5,2)$ bands of
the Meinel system \ by proton impact on N$_{2}$ are presented. Interestingly
enough, the ratio of the cross sections of $(4,1)$ to $(3,0)$ bands, and $%
(5,2)$ to $(3,0)$ bands, are 0.8 and 0.55, respectively. These ratios are
practically independent of the proton energy. The approximate values 0.9 and
0.54 are obtained for these ratios for the case of the electron-N$_{2}$
collision. These results are also in good agreement with previously
published data in Refs. \cite{Malxaz 8} and \cite{Malxaz 22}.

The total excitation cross sections for the N$_{2}^{+}($B$^{2}\Sigma
_{u}^{+})$ and N$_{2}^{+}($A$^{2}\Pi _{u})$ by proton impact on N$_{2}$ are
presented in Fig. 5. For the given states, the total cross sections are
determined by summation of the cross sections that correspond to the
population of the partial vibrational levels. The data analysis for the
Meinel band system shows that the relative population of the separate level
is well described by Franck-Condon principle. Thus, for determination of the
excitation cross section of electronic states, it is sufficient to measure
the excitation function for some intense excitation band.

The behavior of the excitation function for the 1NG N$_{2}^{+}$ $($B$%
^{2}\Sigma _{u}^{+})$ ground state and the excitation of vibrational levels
are different and essentially depend on the energy interval. To determine
the excitation cross sections for the 1NG N$_{2}^{+}$ $($B$^{2}\Sigma
_{u}^{+})$ state the cross sections for the ($0,0$), ($1,2$) and ($2,3$)
excitation bands were measured. Using the decay probabilities of the
vibrational states from Ref. \cite{Malxaz 27}, the excitation cross sections
of the states 0, 1 and 2 were determined. Using the expression

\begin{equation}
\sigma \left( N_{2}^{+}(B^{2}\Sigma _{u}^{+})\right) \approx \overset{2}{%
\underset{\nu =0}{\sum }}\sigma \left( \nu \right)  \label{Eq:2}
\end{equation}%
we determine the total cross section of the excitation state of N$_{2}^{+}$(B%
$^{2}\Sigma _{u}^{+}$)$.$ The total cross section obtained in this way is
quite accurate because the contributions of the excited vibrational states
with $\nu \geq 2$ are small.

To describe collision processes, we use the quasidiatomic approximation,
i.e., the molecule is considered as one centered atomic particle. Therefore
we can use the analogy between the ion-atom and the ion-molecule particles,
so that the data obtained for atoms can be extended to molecules.

\bigskip For charge exchange between the H$^{+}$ and N$_{2}$ molecule, the
intermolecular potential at large distances may be written as:%
\begin{equation}
V_{i}(R)=V_{i}(\infty )-\frac{\alpha _{i}}{2R^{4}},  \label{Eq:3}
\end{equation}%
where $\alpha _{i}$ is the dipole polarizability of the hydrogen atom and
the nitrogen molecule and $i=1,2$ for the reactant and product states,
respectively. The transfer of charge occurs in the region of the critical
distance $R_{c},$ where the coupling-matrix element $H_{12}(R)$ equals half
the difference between the intermolecular potentials:

\begin{equation}
H_{12}(R_{c})=\frac{1}{2}\Delta V_{12}(R)=\frac{1}{2}\left\vert \Delta E-%
\frac{\alpha _{1}}{2R_{c}^{4}}+\frac{\alpha _{2}}{2R_{c}^{4}}\right\vert ,
\label{Eq.4}
\end{equation}%
where $\Delta E=V_{2}(\infty )-V_{1}(\infty )$ is the energy defect for the
charge exchange process. In our case $\Delta E=0.19$, $\alpha
_{1}=11.74a_{0}^{3},$ and $\alpha _{2}=4.5a_{0}^{3},$ where $a_{0\text{ }}$%
is Bohr radius. Following Refs. \cite{Demkov 1, Demkov 2} over the region of
transfer, $R_{c}\pm \Delta R_{c}$, it is a reasonable approximation to use
an exponentially decreasing coupling-matrix element in the simple form:%
\begin{equation}
H_{12}(R)=e^{-\lambda R},  \label{Eq:5}
\end{equation}%
where $\lambda $ is a coupling parameter and all quantities are in atomic
units. The experimental results for the excitation functions for the B$-$%
states are shown in Fig. 6 and are compared with the simple theoretical
estimate \cite{Malxaz 23}

\begin{equation}
\sigma =\frac{1}{2}\pi R_{c}^{2}\sigma ^{\ast }\left( \delta ^{-1}\right) .
\label{Eq:6}
\end{equation}%
In Eq. (6) $\sigma ^{\ast }$ is a tabular function of the parameter $\delta
^{-1}=\hbar \lambda v_{0}/\pi H_{12}(R_{c})$ \cite{Malxaz 23}, where $v_{0}$
is the collision velocity. To calculate the cross section, it is necessary
to estimate the internuclear separation $R_{c}$ at which the transfer
occurs. To find the cross section we need to determine the coupling
parameter $\lambda $ and the value of the matrix element $H_{12}(R_{c}).$ To
determine these values we use self-consistent procedure by varying $R$
within the interval $2.8a_{0}\leq R\leq 4a_{0}$ \ and obtain an appropriate
value of $H_{12}(R_{c})$ from Eq. (4) and then get the corresponding value
of $\lambda $ from (5). This procedure wasconducted in such a way that the
maximum value of the cross section (6) satisfactorily was coincided with the
experimental maximum value in Fig. 6. As a result, we obtained $R_{c}\approx
3a_{0}$, $H_{12}(R_{c})$\ = 0.073 and $\lambda $= 0.87.

The interatomic distance may also be estimated from the expression \cite%
{Malxaz 23}:

\begin{equation}
H_{12}(R)=\sqrt{I_{1}I_{2}}R^{\ast }\text{ }e^{-0.86R^{\ast }},  \label{Eq:7}
\end{equation}%
where $I_{1}$ and $I_{2}$ are the ionization potential of the hydrogen atom
and nitrogen molecule, respectively, and

\begin{equation}
R^{\ast }=\frac{1}{2}(\alpha +\gamma )R.  \label{Eq:8}
\end{equation}%
In Eq. (8) $\frac{1}{2}\alpha ^{2}$ equals the effective ionization
potential $I_{1}$ of the hydrogen atom and $\frac{1}{2}\gamma ^{2}$ equals
the effective ionization potential $I_{2}$ of the nitrogen molecule with all
quantities being in atomic units. For $R=R_{c}$ where the charge transfer
occurs we obtain

\begin{equation}
\sqrt{I_{1}I_{2}}R^{\ast }\text{ }e^{-086R^{\ast }}=\frac{1}{2}\left\vert
\Delta E-\frac{\alpha _{1}}{2R_{c}^{4}}+\frac{\alpha _{2}}{2R_{c}^{4}}%
\right\vert .  \label{Eq:9}
\end{equation}%
Solving Eq. (9) for $R_{c},$ we obtain $R_{c}=3.6a_{0}$ and $%
H_{12}(R_{c})=0.085.$ However, this value significantly shifts the maximum
of the cross section to the region of high energy. In Ref. \cite{Malxaz 29}
it was mentioned that the discrepancy in the estimation for $R_{c}$ can be
explained by the variation of the coupling parameter in the exponent in Eq.
(7) from 0.5 to 2.3.

To determine the value of $R_{c},$ one can also use the table from Ref. \cite%
{Malxaz 23}. From this table it follows that the maximum of the reduced
cross section is $\sigma ^{\ast }=$1.08. By substituting this value in
equation $\sigma ^{\ast }=\sigma /\frac{1}{2}\pi R_{c}^{2},$ where $\sigma
=1.9\times 10^{-16}$cm$^{2}$ is the maximum value of the experimental cross
section in Fig. 6, we obtain $R_{c}\approx 2a_{0}.$ However, this value is
less than the interatomic distance in the nitrogen molecule and, therefore,
it cannot be used since the idea of a polarization interaction loses its
meaning. Probably Eq. (5) for the ion-molecular collision must be corrected
by taking into account Franck-Condon factors.

Based on our results, we can conclude that the population mechanism for B$%
^{2}\Sigma _{u}^{+}$ states in slow collisions is determined by nonadiabatic
transitions between crossing of potential-energy curves, with charge
transfer occurring at the curve crossing (Demkov mechanism \cite{Demkov 1,
Demkov 2}) for the collided particle system. The obtained value of $R_{c}$=
1.6$\times $10$^{-8}$cm corresponds to the distance which is greater than
the size of the molecule. It is obvious that the impact particle collides
with randomly oriented molecular particles. Our data represents results that
are averaged with respect to the orientation. According to our estimation,
we can expect that, the influence of the effect of molecular orientation in
these collision processes is minimal. Otherwise, the maximum of the cross
section will be reduced by at least one order\emph{\ }of magnitude. We can
conclude that, the employed quasimolecular approximation is valid for the
given pairs. All of the above may be extended to the case of excitations of A%
$-$molecular states as well. The maximum of the excitation cross section for
the given states is reached at relatively low collision energy and that is
greater than the maximum of the cross section of B$-$states. This fact
indicates that excitation of the A$-$state due to nonadiabatic transition
occurs at a comparatively large distance and that the value of the
coupling-matrix element $H_{12}(R_{c})$\ is less than in case of the B-state.

\begin{figure}[t]
\centering
\includegraphics[width = 3.5in]{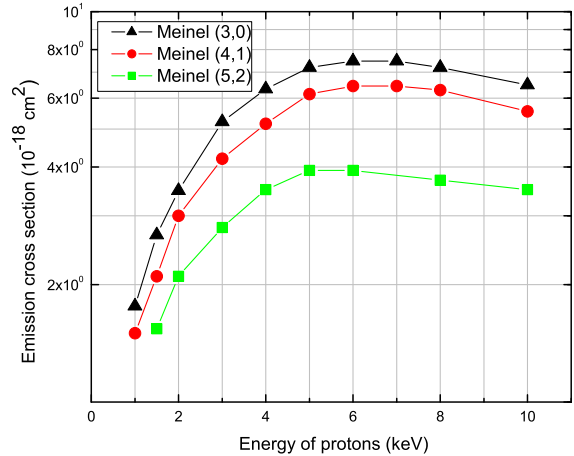}
\caption{Dependence of excitation cross sections for the Meinel (3, 0), (4,
0), and (5, 2) band systems on the proton energy for p$-$N$_{2}$ collisions.}
\label{lam ht4}
\end{figure}

\begin{figure}[t]
\centering
\includegraphics[width = 3.5in]{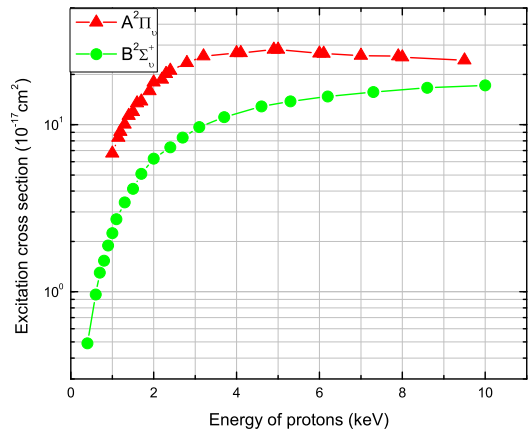}
\caption{Dependence of the manifold excitation cross section for the $A^{2}\Pi _{u}$
and $B^{2}\Sigma _{u}^{+}$ states on the proton energy for p$-$N$_{2}$
collisions. }
\label{lam ht5}
\end{figure}

\begin{figure}[t]
\centering
\includegraphics[width = 3.5in]{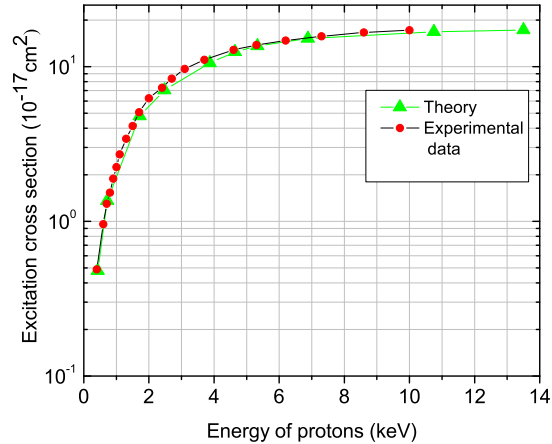}
\caption{Dependence of the excitation cross section for the first negative
band system on the proton energy for p$-$N$_{2}$ collisions.}
\label{lam ht6}
\end{figure}

\section{Conclusion}

\bigskip We presented the measurements of the absolute cross sections for
the $e-$N$_{2}$ and $p-$N$_{2}$ collisions for the first negative $%
B^{2}\Sigma _{u}^{+}-X^{2}\Sigma _{g}^{+}$ and Meinel $A^{2}\Pi
_{u}-X^{2}\Sigma _{g}^{+}$ bands in the energy region 400$-$1500 eV for
electrons and $0.4-10$ keV for protons, respectively. Measurements were
performed in the visible spectral region 400-800 nm by an optical
spectroscopy method. The experimental results were obtained in the condition
when the collision quenching effect was taken into consideration and was
minimized to the single collision condition. This was verified by the linear
dependence of the intensity of spectral lines on the target gas pressure and
the density of the electron current.

In the case of electrons, the measurements of the excitation functions for
the first negative nitrogen molecule ion (B$^{2}\Sigma _{u}^{+}-$X$%
^{2}\Sigma _{g}^{+})$ system with band ($0,0$) and for the Meinel (A$^{2}\Pi
_{u}-$X$^{2}\Sigma _{g}^{+}$) system with bands (3,0), (4,1), and (5,2) show
that the ratio of the cross section of the Meinel band system to the cross
section of the first negative band system $(0,0)$ does not depend on the
energy of incident particles. The population of vibrational levels
corresponding to A$^{2}\Pi _{u}$ states are well described by Franck-Condon
principle.

The ratios of the cross sections of $(4,1)$ to $(3,0)$ bands, and $(5,2)$ to 
$(3,0)$ bands are slightly dependent on the proton energy. The theoretical
estimate within the quasimolecular approximation gives a reasonable
description of the total cross section for the first negative band system of
the nitrogen ion N$_{2}^{+\text{ }}$for $H^{+}-N_{2}$ collisions. This
approach can be extended to the description of excitation of A$-$states as
well, assuming that excitations are due to the nonadiabatic transitions
occurring at large distances. Moreover, it becomes possible to ignore the
orientation of the target molecule for the $H^{+}-N_{2}$ collisions in the
considered energy region of the incident proton.



\end{document}